\documentclass[a4paper,11pt]{article}
\pdfoutput=1 % if your are submitting a pdflatex (i.e. if you have
\usepackage{jinstpub} % for details on the use of the package, please
\title{\boldmath Study of Position Sensitive Silicon Detector (PSD) \\for SiW-ECAL at ILC}

\author[a,1]{Y. Uesugi,\note{Corresponding author.}}
\author[a]{R. Mori,}
\author[a]{H. Yamashiro,}
\author[a]{T. Suehara,}
\author[b]{T. Yoshioka}
\author[a]{and K. Kawagoe.}

\affiliation[a]{Department of Physics, School of Science, Kyushu University}
\affiliation[b]{Research Center for Advanced Particle Physics, Kyushu University,\\744 Motooka, Nishi-ku, Fukuoka, 819-0395 Japan}
\emailAdd{uesugi@epp.phys.kyushu-u.ac.jp}

\abstract{For the past few years we have been developing position sensitive silicon detectors (PSDs) which have an electrode at each of four corners so that incident position of a charged particle can be obtained with signal from the electrodes. It is expected that the position resolution of the silicon-tungsten electromagnetic calorimeter (SiW-ECAL) of the International Large Detector (ILD) will be improved by introducing PSDs to detection layers.
In the previous production we found that the charge separation is not optimal due to the readout impedance. To solve the issue, we produced new PSDs with higher surface resistance with an additional resistive layer on the surface. We also implemented several techniques to decrease position distortion and increase signal-to-noise ratio which are essential for optimal position resolution.
We present first measurements of the performance of one new PSD using a strontium 90 source, and using the Skiroc2-CMS ASIC.
}

\proceeding{Calorimetry for the High Energy Frontier 2019 (CHEF2019)\\
  November 25-29, 2019\\
  Fukuoka, Japan}

\begin{document}
\maketitle
\flushbottom

\section{Introduction}
\label{sec:intro}

The International Linear Collider (ILC) is a future electron-positron collider for precise measurements of Higgs bosons and various BSM searches. A silicon-tungsten electromagnetic calorimeter (SiW-ECAL) is one of the candidates to be used in the International Large Detector (ILD)\cite{a}, one of the detector concepts for the ILC. 
SiW-ECAL has a multilayer structure consisting of 20-30 layers of silicon detectors and tungsten absorbers. The main target of the SiW-ECAL is the measurements of photon energies. %Most of photons from jets come from $\pi^0$ and follow the decay process of  $\pi^0\to2\gamma$. 
Most photons in jets originate from decays of neutral pions. In order to improve the accuracy of $\pi^0$ reconstruction, sensors with high position resolution are desired to precisely measure the direction of photons.
The silicon sensors of the SiW-ECAL are segmented into 5.5 $\times$ 5.5 mm$^2$ cells to maximize the performance of Particle Flow Algorithm (PFA)\cite{PFA}. We are investigating possibility of implementing Position-Sensitive Detector (PSD) technique to each cell of the sensors of the innermost layers of SiW-ECAL in order to improve the position resolution of incident particles, which may lead to improvements on {\it eg.}~PFA performance, $\pi^0$ reconstruction with kinematic fit, and BSM searches with displaced neutral particles.

\begin{figure}[htbp]
\centering
\includegraphics[width=10cm,clip]{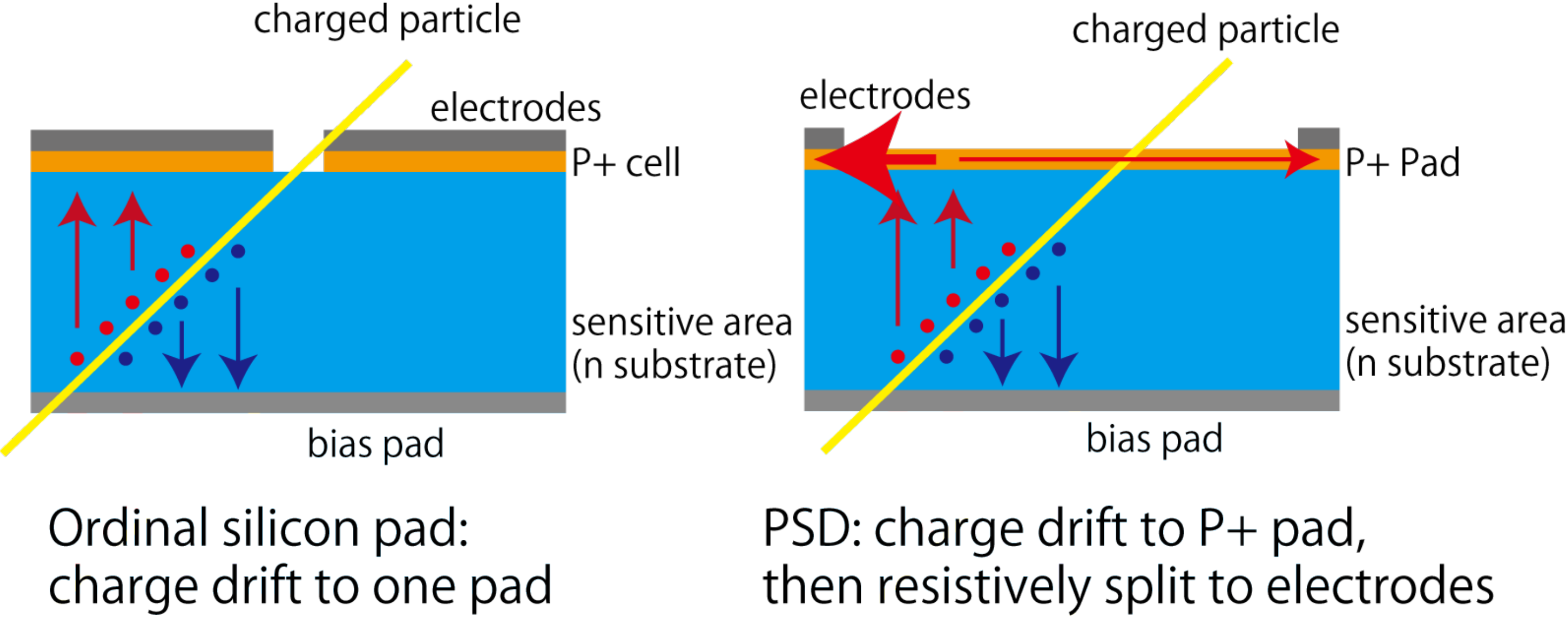}
\caption{\label{fig:psd-e} Comparison of internal structure of silicon sensor and PSD. The left figure shows the traditional silicon sensor used in the SiW-ECAL, and the right figure shows the cross section of the PSD.}
\end{figure}

A schematic view of the simple silicon pad sensors in SiW-ECAL is shown in Fig.~\ref{fig:psd-e} (left). Electron-hole pairs are generated when a charged particle passes through a sensitive area. The reverse bias voltage moves the holes to the P$^+$ pad, and the charge directly goes to the electrodes covering the P$^+$ pad for the readout.
Our PSD is a silicon sensor with segmented cells with similar structure, but each cell has an electrode at each corner instead of a simple pad spread over the cell. When the signal charge reaches a P$^+$ pad, the charge is resistively split to electrodes via a resistive layer on the surface, as shown in Fig.~\ref{fig:psd-e} (right). The hit position is reconstructed as the 2-dimentional center-of-gravity of signal strengths of the electrodes at the four corners. 
In contrast to using smaller cells, the position resolution can be improved with minimal increase of the readout channels if we replace the silicon pads with PSDs in the SiW-ECAL.

In the previous study, photons from a pulsed infrared laser were injected to a single-cell PSD sensor of 7 $\times$ 7 mm$^2$ to demonstrate the position reconstruction\cite{yamashiro}. The position is calculated by
\begin{eqnarray}
  X_{\rm{rec}} &=& \rm{\frac{(Q_0+Q_1)-(Q_2+Q_3)}{Q_0+Q_1+Q_2+Q_3}} 
  \label{eqn:xrec} \\
  Y_{\rm{rec}} &=& \rm{\frac{(Q_0+Q_2)-(Q_1+Q_3)}{Q_0+Q_1+Q_2+Q_3}}
  \label{eqn:yrec} 
\end{eqnarray} \\
where $X_{\rm{rec}}$ and $Y_{\rm{rec}}$ are the reconstruction position in $X$ and $Y$ axes and $Q_i$ is measured charge at each corner electrode. It was expected that $X_{\rm{rec}}$ and $Y_{\rm{rec}}$ should range from $-1$ to $1$, but it appeared that the difference between maximum and minimum $X_{\rm{rec}}$ ($Y_{\rm{rec}}$), called ``dynamic range" in the following discussions, is around 0.3, which degrades the position resolution by a factor 7, compared to the full dynamic range of $1 - (-1) = 2$. The actual position resolution also depends on signal-to-noise ratio.
Distortion of the reconstructed positions compared to the injected positions was also observed. To solve these issues, we developed a new PSD with ideas to increase the dynamic range and reduce the position distortion.

\section{Specifications of the new PSD sensors}

\begin{figure}[htbp]
\centering
\includegraphics[width=11.8cm,clip]{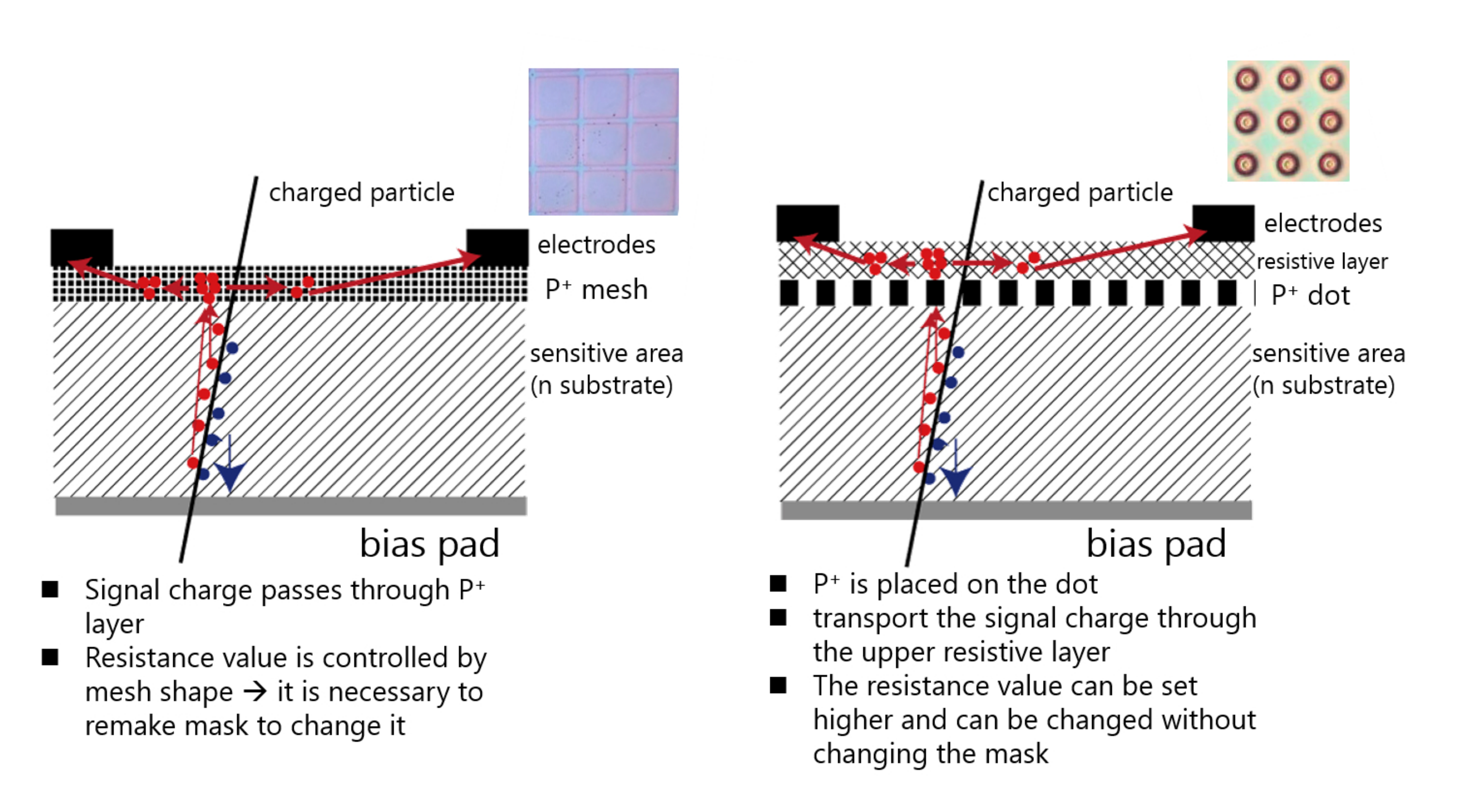}
\caption{\label{fig:resistive_layer}Comparison of two types of surfaces of the PSDs. The left figure shows the meshed P$^+$ surface, and the right figure shows the structure with P$^+$ dots and a resistive layer.}
\end{figure}

It is considered that the small dynamic range is due to the readout impedance, which can be comparable to the surface resistance of the previous PSDs. Since the readout impedance is recognized as serial impedance to the surface resistance, it degrades the dynamic range.
The previous PSD was designed with a meshed P$^+$ surface to increase the resistance from a planar P$^+$ pad as shown in Fig.~\ref{fig:resistive_layer} (left), but it appeared to be insufficient.
Due to technical reasons, higher resistance is difficult with this method. Instead, we adopted a dedicated resistive layer over the P$^+$ layer in the new design to propagate charge with higher resistance as shown in Fig.~\ref{fig:resistive_layer} (right). It also benefits that the resistance can be easily controllable by changing the thickness of the layer, without changing the design of the layer %which costs a lot
reducing costs.
The P$^+$ layer is formed as a matrix of small dots instead of a mesh to prevent charge to spread over the P$^+$ layer. Since the noise can be higher with the larger resistance, %it is necessary to find the optimal resistance to compromise the dynamic range and the noise performance
it is necessary to balance the competing requirements of high dynamic range and low noise by optimizing the resistance.

Among the sensors of the new design, we discuss three types of PSDs in this paper, called PSD1-1, PSD1-2 and PSD2. 
The cell size is $5.5 \times 5.5$ mm$^2$, and the sensor thickness is 650 $\rm{\mu}$m for all the sensors.
It is also common to all the sensors that the left half of the sensors is formed with a meshed P$^+$ surface, as for the previous production, and the right half is formed with a dedicated resistance layer, with the resistance tuned to be 10, 20 and 30 times larger than the meshed P$^+$ surface. The exact resistance value of the PSDs is unknown because the manufacturer does not disclose it, but the resistance between the two electrodes of the meshed P$^+$ surface is around 1 $\rm{k\Omega}$. We produced 6 sensors on each design, with two each having each surface resistance.

\begin{figure}[htbp]
  \begin{tabular}{ll}
     \begin{minipage}[t]{0.5\hsize}
       \centering
       \includegraphics[width=6cm,viewport=6.5cm 18cm 16cm 27.5cm,clip]{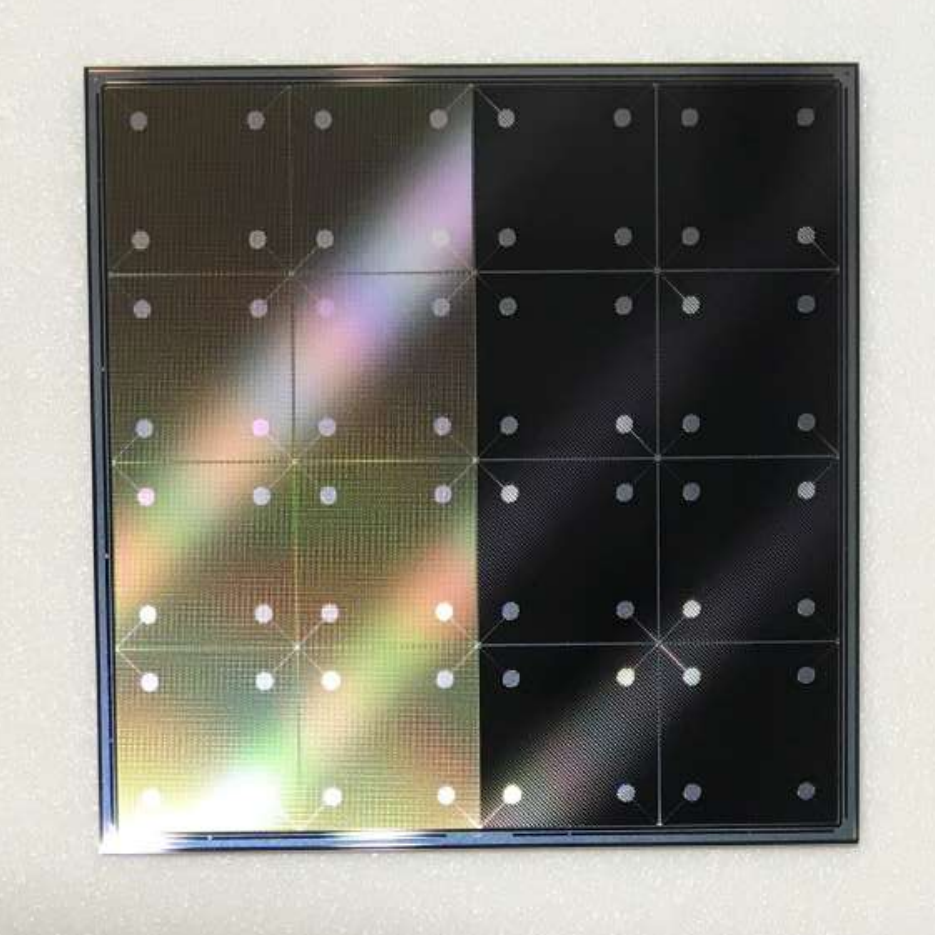}
       \caption{PSD1\label{fig:PSD1photo}}
     \end{minipage}
     \begin{minipage}[t]{0.5\hsize}
        \centering
	\includegraphics[width=6cm,viewport=8.3cm 10cm 15.3cm 17cm,clip]{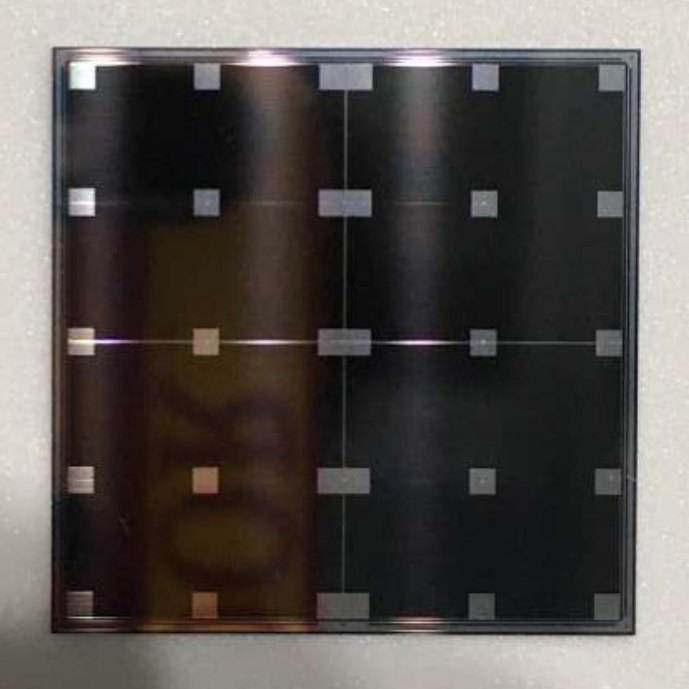}
	\caption{PSD2\label{fig:PSD2photo}}
     \end{minipage}
 \end{tabular}
\end{figure}

\begin{figure}[htbp]
  \begin{tabular}{ll}
     \begin{minipage}[t]{0.5\hsize}
        \centering
        \includegraphics[width=3cm,clip]{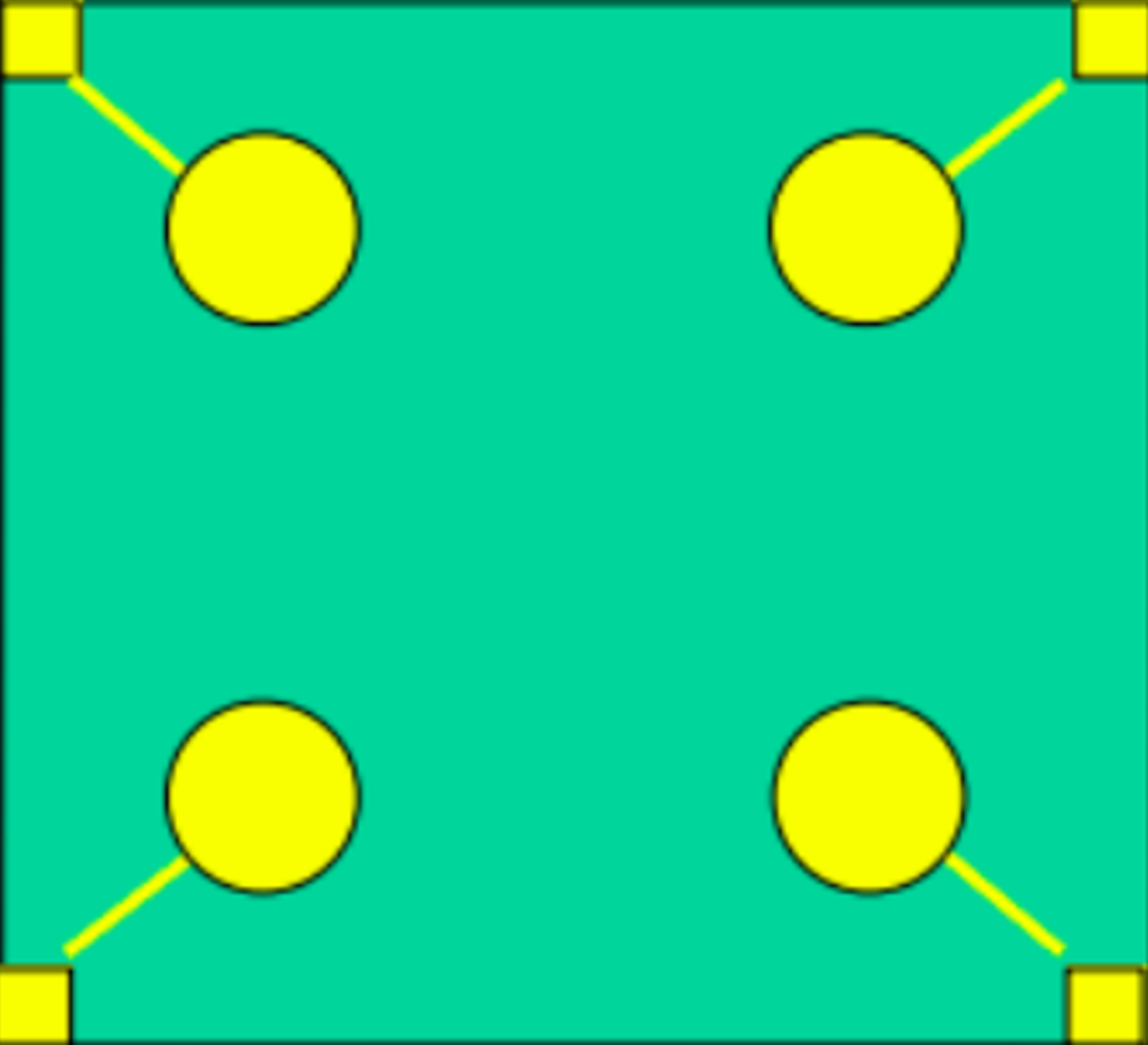}
        \caption{One cell of PSD1\label{fig:PSD1}.}
     \end{minipage}
     \begin{minipage}[t]{0.5\hsize}
       \centering
       \includegraphics[width=2.72cm,clip]{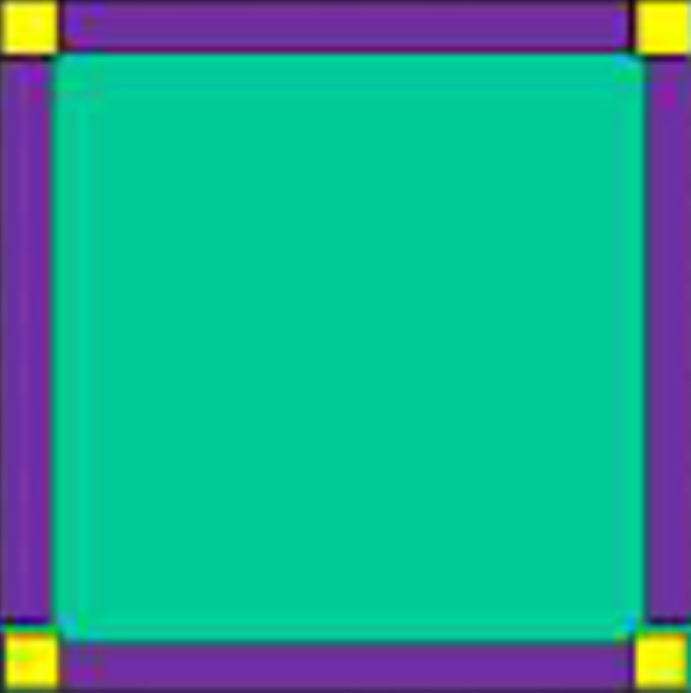}
       \caption{Low resistance lines in one cell of PSD1-2\label{fig:lowresistanceline}.}
     \end{minipage}
  \end{tabular}
\end{figure}

Figures \ref{fig:PSD1photo} and \ref{fig:PSD2photo} are photographs of PSD1 (PSD1-1 and PSD1-2) and PSD2, respectively. PSD1-1 and PSD1-2 has the same patterns for the electrodes. The schematic view of one pixel of PSD1 is shown in Fig.~\ref{fig:PSD1}. 
Circular pads are connected to the electrodes at the four corners to avoid large electrodes at the corners preventing the laser injection on the corner region. PSD1 has 4 by 4 cells, with 4 channels per cell, having a total of 64 channels of the readout electrodes.

The difference between PSD1-1 and PSD1-2 is the resistance pattern.
PSD1-1 is formed with flat surface resistance over each cell.  
In contrast, PSD 1-2 has low resistance lines at the cell edges to reduce distortion, which is a proven method discussed in \cite{c}.
The schematic view of the resistance pattern is shown in Fig.~\ref{fig:lowresistanceline}. 
The edge resistance is 4 (on the upper cells of the sensors) and 8 (on the lower cells of the sensors) times smaller than the resistance over the surface.


The PSD2 features the sharing of an electrode with neighbor cells, as shown in Fig~\ref{fig:PSD2photo}, to suppress an increase of the number of the readout channels. This enables the realization of a PSD with a similar cell size to non-PSD silicon pad sensors keeping the number of the readout channels the same.
For example, a non-PSD silicon sensor with 256 readout channels can have 16 by 16 cells. If we replace it by a PSD with the same number of readout channels, we have to reduce the number of cells to 8 by 8 with PSD1 structure, while with PSD2 we can have 15 by 15 cells.

\section{$\beta$ radiation source measurement}
Radiation source measurements were not available at the time of the CHEF conference but, as they are of interest to the community, are presented here.
In this study, we acquire the PSD signals using an ASIC called Skiroc2-CMS and its evaluation board, developed by the Omega / IN2P3 group. A Skiroc2-CMS has 64 readout channels for silicon sensors, which is suitable for the readout of one PSD sensors. The Skiroc2-CMS ASIC\cite{skiroc2cms} is made for silicon sensor cells with up to 70 pF detector capacitance per channel, which meets our PSD sensors having $\sim$40 pF per cell. After a preamplifier, it has three shapers; two slow shapers with a 40 ns shaping time for 12-bit ADCs with high and low gain and a fast shaper with a few ns shaping time for a timing trigger. We performed all measurements in this section with self triggering using this timing trigger. We only used ADC output with the high gain shaper in this study.
Input pins of the Skiroc2-CMS are connected to a sensor board via a 80-pin connector (HRS FX10A-80 series). PSD sensors are connected to the sensor board by conductive glue (EPO-TEK E4110-LV) using a glue-dispensing robot and covered with the sensor board and black tape.
To apply high voltage (HV) for the sensor bias, a HV board is connected to the sensor by a conductive adhesive tape. 150 V is supplied as the bias voltage from a voltage source, with a low-pass filter on the HV board for noise reduction.

A $^{90}\mathrm{Sr}$ source of 10 kBq is used for the irradiation measurement. The source is placed at about 12 mm distance from the PSD surface. The source is located near the center of the PSD so that the source illuminate the whole PSD almost uniformly. Since $^{90}\mathrm{Sr}$ is a pure electron source, we expect signal similar to minimum ionizing particles (MIPs), giving the avarage of  7.8 fC charge deposit with 650 $\mu$m silicon sensor, assuming full depletion.
We checked the difference of trigger frequency with and without the source. The frequency is more than 1000 times higher with the source, which confirmed the signal with irradiation. The width of the pedestal is measured to be 4-5 ADC count, which corresponds to $\sim$2.5 \% MIP or 0.19 fC.

\begin{figure}[htbp]
 \centering
 \includegraphics[width=4.5cm,viewport=7cm 0.3cm 12.78cm 13.5cm]{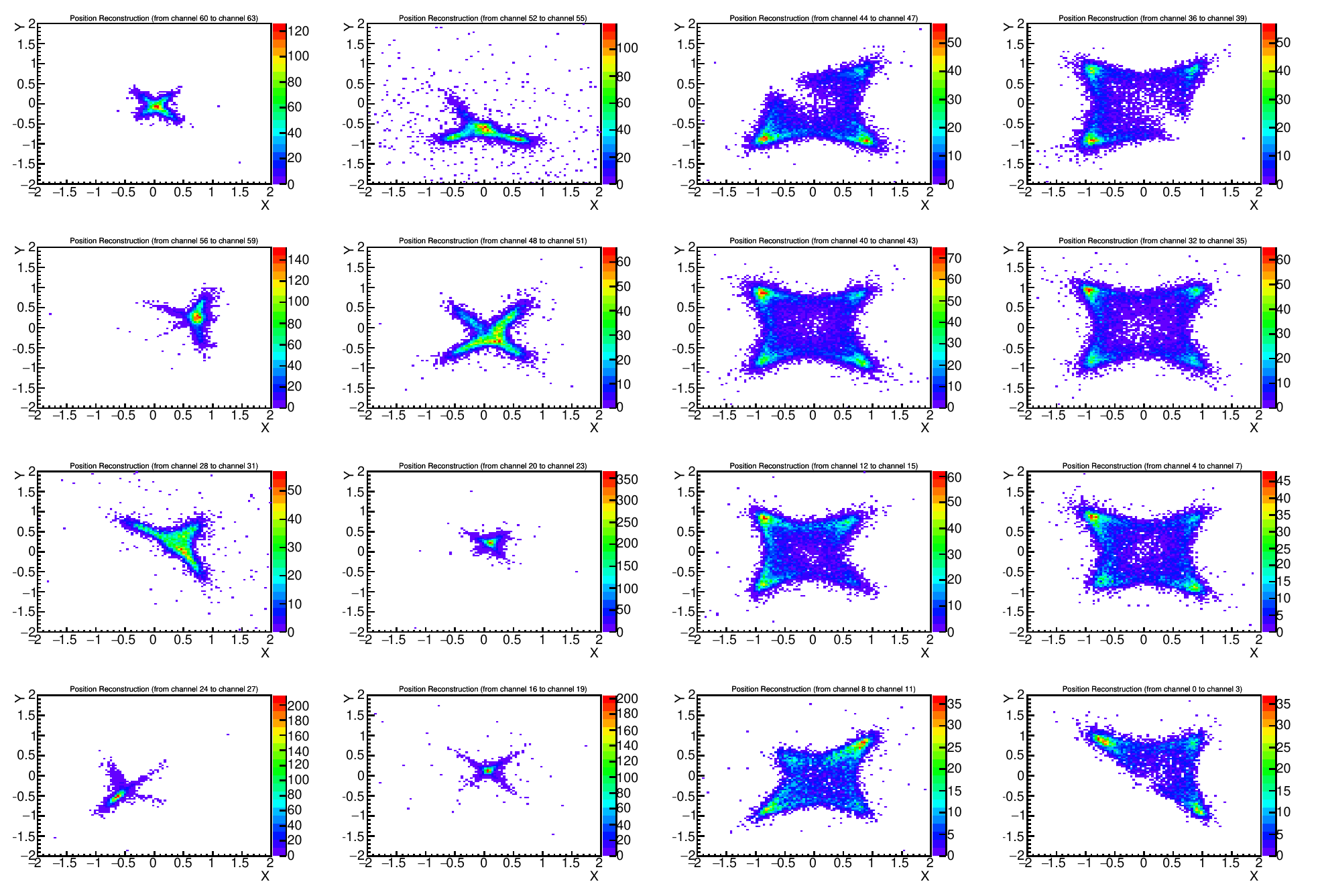}
 \caption{\label{fig:farPSD1-2-2sr90}
2D-distribution of reconstructed positions with PSD1-2 using a $^{90}\mathrm{Sr}$ soruce.
Triggers on the left-top electrode of the first row, the third column and the bottom-right electrode of the first row, the fourth column are disabled due to higher noise than other channels.}
\end{figure}

The hit position is reconstructed using equations \ref{eqn:xrec} and \ref{eqn:yrec} after pedestal subtraction. Fig.~\ref{fig:farPSD1-2-2sr90} shows the distributions of reconstructed hit positions %at each cell with irradiation
for the 16 irradiated cells in PSD1-2. %PSD 1-2 with the resistance layer 10 times larger resistance from the meshed P$^+$ surface is used for the measurement.
The expected position ranges from $-1$ to $1$, while the distribution is shown as the range of $-2$ to $2$ to cover fluctuations due to random noise.
As already discussed, the left half is equipped with the meshed P$^+$ with lower resistance. It shows that most of the signal is concentrated on a small region of the cells on the left half. This is due to the relatively low dynamic range as expected.
In contrast, the right half gives wider distribution almost reaching the full dynamic range of $-1$ to $1$, which shows that the surface resistance of the cells is high enough. However, it shows several issues. First, we see the concentration of the hits at four corners of each cell. This is considered to be related to the self-triggering feature. Although the trigger should be optimally decided with a sum of four channels connected to one PSD cell, it is done with a threshold on single channel due to the limitation of the ASIC function.
Due to rather discrete steps of the threshold control, we had to set the threshold around half of the MIP charge. 
Considering that the signal is split to four electrodes especially for hits around the center of a cell,
this causes significant inefficiency of the trigger in the center region, causing more events on the corner and edge region.
Since it was found that the detection efficiency of the self triggers was low, the use of an external trigger counter will also be considered.
Another problem is that we still see distortion of the distribution regardless of the low resistance lines implemented on the edges of each cell.
We do not see significant difference between two specifications of the edge resistance (by comparing upper and lower half).
This should be investigated with measurements with other PSDs, which is ongoing.
We also plan to confirm the non-flat response by measurements with laser injection.

\section{Summary}
We developed a new series of PSDs with several ideas to improve the dynamic range and the position distortion.
Wider dynamic range is observed with beta irradiation on a PSD with larger surface resistance.
We need further studies of laser measurements and radiation measurements for various specifications of PSDs such as PSD2 and comparison between PSD1-1 and PSD1-2.
We will make Spice simulations for better understanding and further optimization.
After optimization of the structure of the PSDs, we plan to develop prototype sensors to be used for the ILD SiW-ECAL.

\acknowledgments

%This is the most common positions for acknowledgments. A macro is available to maintain the same layout and spelling of the heading.

We appreciate Omega group for the support on the operation of the Skiroc2-CMS chip.
This work is partially supported by JSPS KAKENHI Grant Number JP17H05407.

%\paragraph{Note added.} This is also a good position for notes added after the paper has been written.



\begin{thebibliography}{99}

\bibitem{a}
T. Behnke, James E. Brau, Philip N. Burrows, M. Peskin, et al., \emph{The International Linear Collider Technical Design Report - Volume 4: Detectors}, %\emph{J. Abbrev.} {\bf vol} 
(2013), arXiv:1306.6329.

\bibitem{PFA}
M.A. Thomson, \emph{Particle Flow Calorimetry and the PandoraPFA Algorithm}, Nucl. Instrum.
and Meth, vol. A611, pp. 25-40, (2009).

\bibitem{yamashiro}
H. Yamashiro, K. Kawagoe, T. Suehara, T. Yoshioka, S. Yuji, and H. Sumida, \emph{Performance evaluation of PSD for silicon ECAL}, in \emph{Proceedings of International Workshop on Future Linear Colliders 2016 (LCWS2016)}:
Morioka, Iwate, Japan, December 05-09, 2016, 2017. [Online]. Available: https://inspirehep.net/record/1518936/files/arXiv:1703.08091.pdf.

\bibitem{c}
A. Banu, Y. Li, M. McCleskey, M. Bullough, S. Walsh, C. A. Gagliardi,
L. Trache, R. E. Tribble, and C. Wilburn, \emph{Performance evaluation of
novel square-bordered position-sensitive silicon detectors with four-corner
readout},
Nucl. Instrum. and Meth, vol. A593, pp. 399-406, (2008).

\bibitem{skiroc2cms}
J. Borg, S. Callier, D. Coko, F. Dulucq, C. de La Taille, L. Raux, T. Sculac 
and D. Thienpont, \emph{SKIROC2\_CMS an ASIC for testing CMS HGCAL}, Journal of Instrumentation, vol. 12, (2017), presented at the \emph{Topical Workshop on Electronics for Particle Physics (TWEPP2016)}, 26-30 September 2016. 

% Please avoid comments such as "For a review'', "For some examples",
% "and references therein" or move them in the text. In general,
% please leave only references in the bibliography and move all
% accessory text in footnotes.

% Also, please have only one work for each \bibitem.


\end{thebibliography}
\end{document}